# Magnetoelectric coupling in epitaxial orthorhombic YMnO$_3$ thin films

X. Martí[1], V. Skumryev[2,3], V. Laukhin[1,2], F. Sánchez[1],

M.V. García-Cuenca[4], C. Ferrater[4], M. Varela[4], and J. Fontcuberta[1]

[1] Institut de Ciència de Materials de Barcelona-CSIC, Campus UAB, Bellaterra 08193, Spain

[2] Institut Català de Recerca i Estudis Avançats (ICREA), Barcelona, Spain

[3] Departament de Física, Universitat Autònoma de Barcelona, Bellaterra 08193, Spain

[4] Departament de Física Aplicada i Òptica, Universitat de Barcelona, Diagonal 647, Barcelona 08028, Spain

## Abstract

We have grown epitaxial thin-films of the orthorhombic phase of YMnO$_3$ oxide on Nb:SrTiO$_3$ (001) substrates and their structure, magnetic and dielectric response have been measured. We have found that a substrate-induced strain produces an in-plane compression of the YMnO$_3$ unit cell. The temperature-dependent magnetization curves display a significant ZFC-FC hysteresis at temperatures below the Néel temperature (T$_N \approx$ 40K). The dielectric constant increases gradually (up to 26%) below T$_N$ and mimics the ZFC magnetization curve. We argue that these effects are a manifestation of magnetoelectric coupling in thin films and that the magnetic structure of YMnO$_3$ can be controlled by substrate selection.

**Introduction**

Understanding and tailoring coupling between the magnetic and dielectric properties of oxides is a very active field of research. The ultimate goal, being to control the magnetic (dielectric) state by using an electric (magnetic) field, requires that both physical properties are coupled. Manganese oxides $AMnO_3$, where A is a suitable rare-earth, present an hexagonal or perovskite (orthorhombic) structure depending on the size of the rare-earth cation; for the smaller (Ho-Lu, Y) rare-earths the hexagonal phase (in bulk) is the equilibrium phase, whereas the orthorhombic phase is obtained for the larger rare-earth (Tb, Dy, Gd). Of relevance here is that whereas at low temperature both families of compounds display an antiferromagnetic (AF) order, the ferroelectric (FE) order sets in at temperatures ($T_C^{FE}$) much above the AF Néel temperature ($T_N$) in the hexagonal structure but it occurs at temperatures below $T_N$ in the orthorhombic ones. For instance: for the hexagonal $YMnO_3$ $T_C^{FE} \sim 900K$ and $T_N \sim 80$ K [1] and for the orthorhombic $TbMnO_3$ $T_C^{FE} \sim 30K$ and $T_N \sim 40$ K [2].

The magnetism and ferroelectricity usually excludes each other and thus the occurrence of biferroicity in these oxides has been disputed. Even more intriguing is the existence of some coupling between these ferroic orders. It appears that FE in hexagonal manganites is associated to tilting of the Mn-O octahedral [3] whereas in the orthorhombic $TbMnO_3$, it has been proposed that it may result from the existence of magnetic transition below $T_N$ –from sinusoidal to helical spin order, that breaks the spatial inversion symmetry thus allowing the existence of ferroelectricity [4].

$YMnO_3$ is at the verge of stability between hexagonal and orthorhombic phases. Although under normal synthesis conditions, the hexagonal phase is the stable one, the metastable orthorhombic o-$YMnO_3$ phase can be obtained using either high pressures (bulk materials) [5] or appropriate epitaxial strain engineering (thin films) [6, 7, 8]. It has been reported that polycrystalline o-$YMnO_3$ display a pronounced dielectric constant ($\varepsilon$) anomaly at $T_\varepsilon < T_N$ [9] which has been taken as a signature of magnetoelectric coupling. The AF magnetic structure of o-$YMnO_3$ has been studied in great detail [10, 11, 12]. Muñoz et al. concluded that the $Mn^{3+}$ spins form a sinusoidally modulated structure (characterized by fluctuating average magnetic moments) defined by a ($C_x$,0,0) mode with the magnetic moments oriented parallel to the propagation vector $\boldsymbol{k}$ =

$(k_x,0,0)$ [11]. The structure was found to remind stable and incommensurable down to 1.7 K and the propagation vector to be blocked at a constant value below about 28 K.

The existence of sinusoidal magnetic structure suggests the occurrence a subtle equilibrium among different magnetic interactions, which are super-exchange and thus should be strongly dependent on the Mn-O-Mn bond topology (angles and distances). Therefore one could expect that epitaxial strain in o-YMnO$_3$ thin films may affect severely the precise nature of the magnetic ordering thus allowing tuning the magnetoelectric response.

In this manuscript we shall address some of these issues. We report first on the stabilization and detailed structural analysis of the epitaxial o-YMnO$_3$ films on suitable substrates. Next we shall provide magnetic and dielectric measurements on these films and we will show that, -for the first time in thin films and in good agreement with earlier reports for polycrystalline materials- the films are AF below $T_N \approx$ 40 K and display a well defined anomaly in $\varepsilon(T)$ at $T<T_N$. Interestingly enough, the magnetic data shows some irreversibility that indicates the presence of a ferromagnetic component. These results suggest that indeed, substrate-induced strain can be used to modulate the magnetic structure and eventually the dielectric response. Unfortunately, under application of an external magnetic field we didn't find a clear evidence of magnetocapacitance.

**Experimental**

YMO$_3$ (YMO) thin films were deposited on (001) oriented Nb:SrTiO$_3$ (NbSTO) substrates (0.5 mm thick) by pulsed laser deposition. A KrF excimer laser (248 nm wavelength, 34 ns pulse duration) was used at a repetition rate of 5 Hz. The laser beam was focused to a fluence of 1.4 J/cm$^2$ on a stoichiometric YMnO$_3$ target, being the substrate placed at a distance of 5 cm. The films were deposited at a substrate temperature of 800 ºC in a 0.2 mbar of oxygen pressure. At the end of the growth, the samples were cooled down and 1 atm of oxygen was introduced in the chamber at 530 ºC. The growth rate is ~0.10 Å/pulse; the thickness of the films was determined from low angle X-ray reflectometry measurements using Cu K$_\alpha$ radiation. The YMO film here reported is 150 nm thick. The crystal structure, epitaxial relationships and lattice strain (out-of-plane and in-plane) were investigated by a Gadds-8 X-ray diffractometer by Bruker. $\theta/2\theta$ and $\omega$-scans of symmetrical reflections were performed, as well as $\phi$-scans and pole figures of asymmetrical reflections.

Magnetic properties were measured by using a superconducting quantum interference device system. Dielectric properties were measured, inside a Physical Property Measuring System (PPMS) from Quantum Design, by using a 16047A impedance analyzer by Hewlett-Packard. The electrical contacts on the YMO/NbSTO with area 8 mm$^2$ were made (top and bottom) by using graphite paste. The background impedance of the wiring and substrate was determined by measuring under nominally identical conditions, the impedance of a bare NbSTO substrate.

**Results and discussion**

X-ray diffraction experiments have confirmed that the YMO film grown on the Nb:STO(001) substrate is orthorhombic. The θ-2θ scan (Figure 1) shows the substrate reflections and the (002) and the (004) peaks corresponding to the orthorhombic phase of YMO. There are no traces of neither other reflections of the orthorhombic phase nor the hexagonal phase. Thus, within the experimental resolution, the film is purely orthorhombic and it is (00l) textured, using the P*bnm* setting ($a$ = 5.26 Å, $b$ = 5.85 Å, $c$ = 7.36 Å in bulk). This is in agreement with previous results [6,7,8] reporting the stabilization of this phase in films epitaxially grown on STO substrates. The out-of-plane lattice parameter of the film has been determined from the position of the YMO(004) peak, after a calibration by using the STO(002) reflection. The value, $c_{film}$ = 7.43 Å, indicates an expansion of the lattice parameter ($c_{bulk}$ = 7.36 Å, thus the strain being $\varepsilon_{[001]}$ = +0.95%). On the other hand, we show in the inset of Figure 1 the rocking curve ($\omega$ scan) of the YMO(004) reflexion, which has a full-width at half maximum of 0.55º whereas the instrumental resolution is 0.02º as determined by rocking curves on STO(002) peak.

Although the rocking curve of this symmetrical reflection is only relatively narrow, the film is epitaxial as evidenced by poles figures. In Figure 2a we have collected the poles from the YMO(111) and the STO(111) planes. As expected, there are four narrow high intensity (111) peaks, 90º apart, which arise from the substrate. In contrast, the four peaks corresponding to the (111) reflexions from the YMO film are notably wide (around 15º) along ϕ. The four peaks are, as the substrate ones, roughly ~ 90º apart, and rotated in-plane by 45º, with respect those of STO(111) planes. We notice that from the orthorhombic symmetry of YMO, two pairs of peaks were expected, with angular splitting Δϕ differing gradually from Δϕ = 90º as higher is the orthorhombicity ($b/a$). Therefore a detailed analysis of the YMO(111) reflections of the pole-figure in Fig. 2a is required. An intensity profile of one of the YMO(111) peaks of Fig. 2a is shown in Fig. 2b. It becomes clear that in fact, each spot is composed by two peaks, so each reflection is doubled. Thus, there are eight peaks that correspond to a pair of crystal variants, with orthorhombic symmetry and 90º rotated in-plane. Taking into account the position of the STO(111) reflections, epitaxial relationships [100]YMO(001) / [110]STO(001) and [010]YMO(001) / [110]STO(001) are obtained. It has to be noted that if the film were fully relaxed with in-plane cell parameters identical to those of bulk YMO, the two peaks in Figure 2b should be separated by 6.06º. However, they are separated by 5.76º, and this means that the film is not fully relaxed but partially strained. Similar films grown on undoped STO(001) substrates, showed the same splitting in $\phi$ and the same out-of-plane lattice parameter, and detailed characterization by means of reciprocal space maps allowed determination of the in-plane lattice parameters. The in-plane cell parameters of the (001) o-YMO film were found to be $a = 5.26$ Å and $b = 5.80$ Å; this indicates that the film is fully relaxed along the [100] crystal direction but only partially strained along the [010] direction ($\varepsilon_{[010]} = -0.86\%$). Additional characterization of the twinning and a discussion of its impact on the relief of the film's elastic energy will be reported elsewhere [13]. Notice that the Mn-spins in the o-YMO lie along the partially strained [010] direction. Therefore, the interatomic Mn-O-Mn bonds in the YMO film are expected to vary with respect to the bulk ones, subsequently modifying the magnetic properties.

The temperature dependence of the magnetization of the YMO/NbSTO sample was measured under different in-plane magnetic field after zero-field-cooled (ZFC) and field-cooled (FC) conditions. The open and closed symbols in Fig. 3a correspond to the FC and ZFC data collected using 1kOe magnetic field. Inspection of the data collected shows that the ZFC-FC magnetization curves split apart at temperatures below ~ 45 K, that we identify with $T_N$; the ZFC magnetization curve displays a broad maximum at around ~ 25 K at decreases gradually at lower temperature. In contrasts the FC magnetization curve keeps growing when further decreasing temperature.

Before discussing the magnetic data it is worth to recall that in these experiments the magnetic field is applied in-plane and thus, according to the information extracted from the structural analysis, the field is parallel to the *a-b* plane which is the plane where, in bulk materials, the Mn magnetic moments are confined. Under these experimental conditions, the magnetization of an ideal AF material, would display a kink at the Néel temperature followed by some non-hysteretic magnetization reduction at still lower temperature. Certainly the data in Fig. 3a does not follow these trends.

The absence of sharp features in the dc-magnetization curves, at the Néel temperature of o-YMnO$_3$ has been signalled earlier although its ultimate reason has not been definitely settled and somehow contradictory results have been reported; indeed in the dc-magnetization data reported by Muñoz et al [11], a only a subtle kink is visible at about 40 K. In contrast, earlier measurements by Wood et al [5] and more recent data Lorentz et al [9] display a lambda-like peak at $T_N$. The observation of hysteresis ZFC-FC is interesting by its own. Hysteretic behaviour is also visible in the data reported by Lorentz et al [9]. Muñoz et al also noticed [11] that the ac-magnetic susceptibility displays a maximum at around T = 11 K which they attributed to the freezing temperature associated to a partial-spin glass behaviour of the sample, although no magnetic hysteresis was observed. It was argued that this spin-glass behaviour could be originated by the presence of some Mn$^{4+}$ ions originated from tinny oxygen excess, and the subsequent magnetic disorder. Competing interactions in this system can be easily introduced if due to some non-stoichiometry related defects, Mn$^{3+/4+}$ states are formed. Double-exchange like Mn$^{3+/4+}$ - Mn$^{3+/4+}$ ferromagnetic interactions can be introduced, thus leading to competing interactions or even ferromagnetic clusters that may behave as a superparamagnet and eventually a spin-glass like behaviour. As, divergence of ZFC-FC curves is found both in spin-glass systems and in superparamagnetic systems, this observation cannot be taken has a unique hallmark of spin-glass behaviour.

We also notice that hysteretic ZFC-FC could also be expected if there is a magnetic canting that give rise to a long range ordered weak-ferromagnetism (w-FM). The present experiments do not allow discriminating among these different scenarios. As already reported in the literature, linear susceptibility may not be sensitive enough to unravel the origin of magnetic hysteresis but higher-order harmonic susceptibility analysis is required [14]. Indeed this is a crucial point that would require further experimental efforts.

May be of higher relevance here is that the strain induced by the substrate, and the resulting change of bond angles and distances, must modify some magnetic interactions eventually reinforcing the competition between them. This is a very appealing possibility and the fact that the ZFC-FC divergence is much pronounced in these o-YMnO$_3$ thin films that what was reported earlier for bulk materials suggest that the indeed substrate induced strain can play an important role.

Now, we address the dielectric measurements performed on the very same sample. Measurements of permeability and magnetodielectric response in thin films are particularly challenging [15]. Therefore we shall discus in some detail the employed methodology. Two samples: a bare NbSTO substrate and the o-YMO/NbSTO film-substrate, were installed in the sample holder of the PPMS systems and, the after contacting, their impedance was measured using two dedicated wires.

Figures 4a and 4b show the temperature dependence of the capacity ($C_s$) of a bare NbSTO substrate (open symbols) and the ($C_{sf}$) o-YMO/NbSTO structure (closed symbols). These data are extracted from raw complex impedance measurements using a excitation voltage of 1V at 100 kHz; the measured total impedance and conductance are of about 500 $\Omega$ and 1.5 mS respectively. The capacitance C is subsequently calculated (displayed in Figures 4a and 4b) assuming two equivalent resistor-capacitance (RC) circuits: a series RC (panel a) and parallel RC (panel b). We note in Figure 4 that both model-circuits lead to the values of capacity very similar. In addition, in the frequency range between 1 kHz and 1 MHz the magnitude of the capacity is also constant (see inset in panel b), therefore we assumed that the major contribution is due to the capacity of the sample under study a free of spurious effects.

In Figures 4a and 4b we observe that the contribution from the bare substrate $C_s$ (open symbols) is monotonously decreasing as the sample is cooled down. The behaviour of $C_{sf}$, the o-YMO/NbSTO films-substrate sample (closed symbols) is very different. Whereas $C_S$ decreases monotonically as the system is cooled down, the $C_{sf}$ contribution exhibits a peak with the maximum value at T = 16 K. It is remarkable that this $C_{sf}$ peak start to emerge at the same temperature where the magnetic order is established in the o-YMO film as we deduced from the magnetic measurements presented above. In order to extract the electric permittivity of the o-YMO film, we proceeded as follows. We modelled the data of Figure 4a with a simple circuit: a series combination of capacitors: $C_{sf}^{-1} = C_b^{-1} + C_f^{-1}$ where the term "$C_b$" includes all parasitic contributions due to the substrate and wiring. Next, we solve for $C_f$ and extract the permittivity of the o-YMO film considering an ideal parallel capacitor with parallel electrodes, $C_{FILM} = \varepsilon \cdot A/t$, where A and t (A= 8 mm$^2$ and t = 150 nm) stand for the area of the graphite contacts and the thickness, respectively. The temperature dependence of the dielectric constant, measured increasing T or decreasing T (open and closed circles respectively) is shown in Figure 3b. We first note that the magnitude of $\varepsilon \approx 20\text{-}25$ is very similar to that reported by Lorentz et al [11] polycrystalline samples. Next, it is clear that when cooling down across the AF transition ($\sim$ 45 K) $\varepsilon$(T) starts growing and reaches a maximum (about 26 % larger) at about T = 16 K. Notice that the temperature dependence of the dielectric permittivity mimics the magnetization curve. Thus the observed behaviour is just the counterpart of the magnetodielectric effect observed by Lorentz et al [11] in bulk o-YMnO$_3$. The subtle thermal hysteresis visible in Fig. 2b is much smaller than that observed in that polycrystalline samples. Magnetocapacitance could be also visible when applying an external magnetic field. Thus we measured the capacitance of the films under a magnetic field of 9 T. Unfortunately, any change in capacitance, if exist, is below the noise level.

**Summary**

In summary, we have shown that epitaxial orthorhombic YMnO$_3$ films grown on SrTiO$_3$ substrates, display a remarkable change of dielectric permittivity when the antiferromagnetic order stets-in. This observation mimics previous results reported for bulk materials. Moreover, we have observed a genuine ZFC-FC hysteresis in the magnetization curves, not occurring in bulk materials, suggesting a strain-induced modification of the magnetic interactions. We have proposed that this behaviour could be interpreted as due to the appearance of long range ferromagnetic order associated to the occurrence of a strain-induced weak ferromagnetism.

**Acknowledgements**

Financial support by the MEC of the Spanish Government (projects NAN2004-9094-C03 and MAT2005-5656-C04) and by the European Union (project MaCoMuFi (FP6-03321) and FEDER) are acknowledged.

**Captions**

Figure 1: XRD diffraction pattern of the YMO/NbSTO(001) film showing only peaks corresponding to YMnO₃ orthorhombic phase. Inset: ω-scan around the YMO(004) reflection.

Figure 2: (a) XRD pole figure collecting the STO(111) and the YMO(111) reflections; (b) Intensity profile along ϕ corresponding to a YMO(111) reflection showing the double contribution in this reflection.

Fig. 3: The plot shows (a) the ZFC – closed symbols - and the FC – open symbols - temperature dependence of the magnetization and (b) capacitance on the very same YMO/NbSTO(001) sample. Later measurements were performed cooling (closed symbols) and warming (open symbols).

Fig. 4: Detail on the extraction of the values for permittivity. The plots show the capacity raw data obtained using either (a) a parallel resistor or (b) a series resistor models. Open symbols correspond to a bare Nb-doped STO substrate whereas the closed symbols correspond to the YMO/NbSTO sample.

**Figure 1**

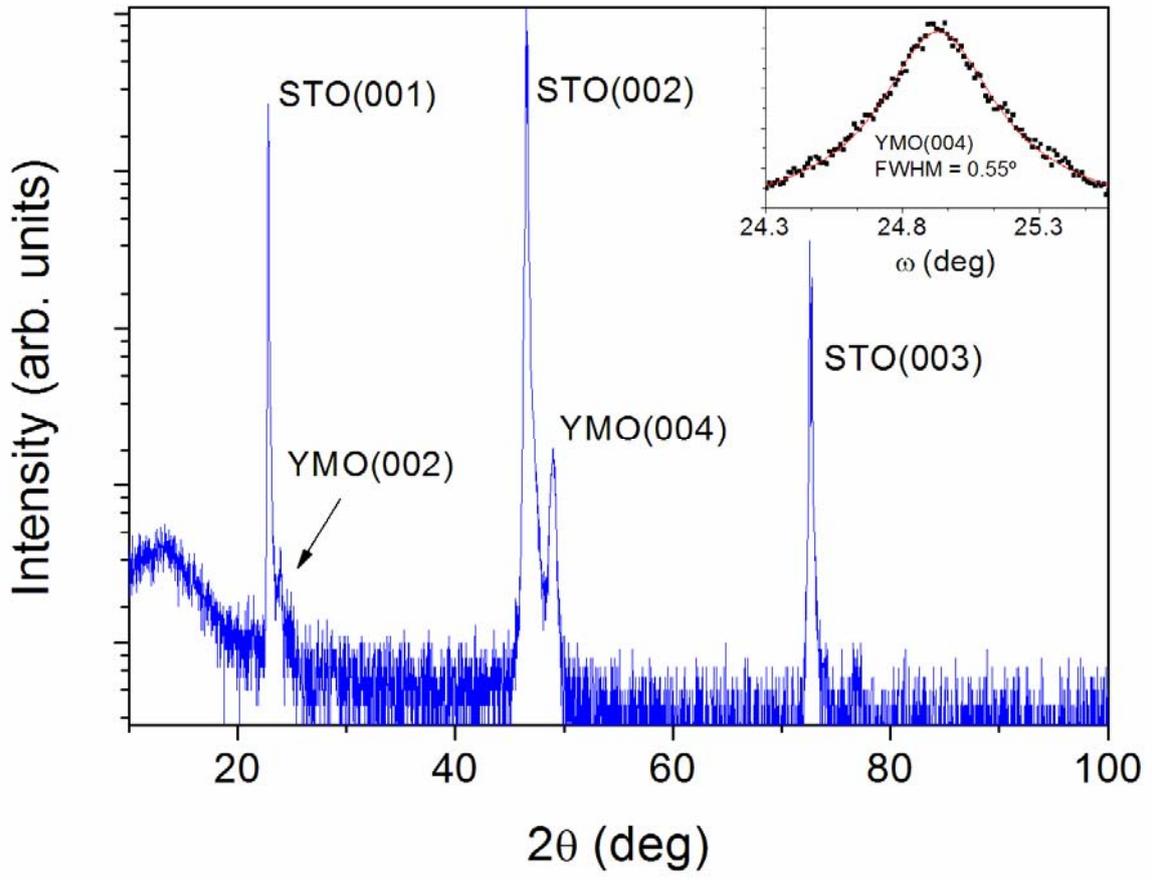

**Figure 2**

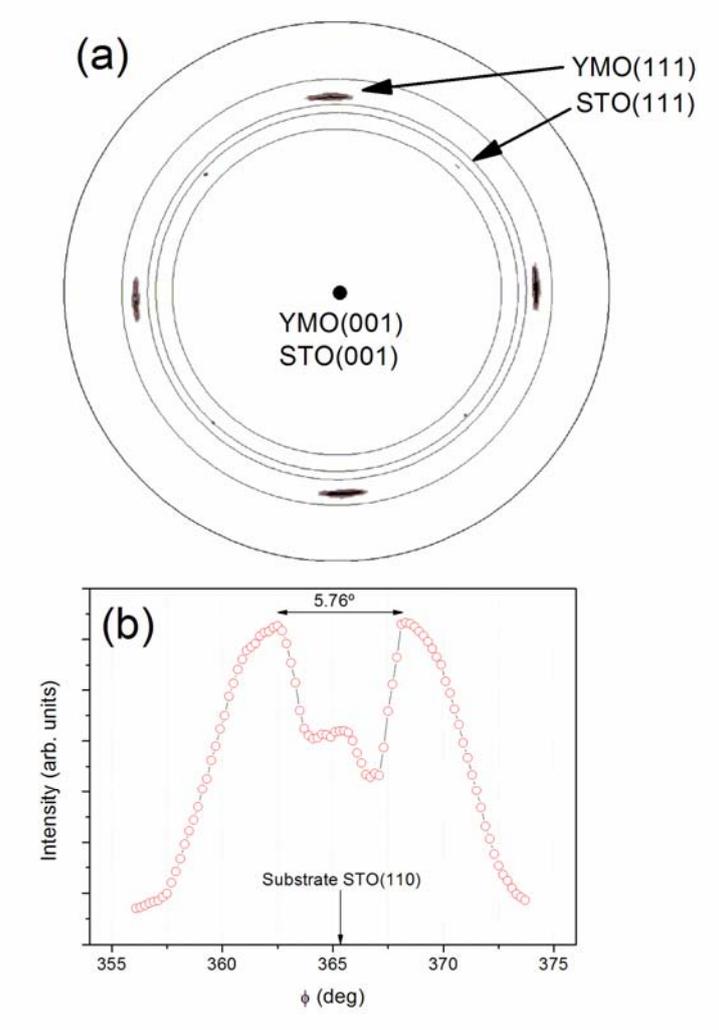

**Figure 3**

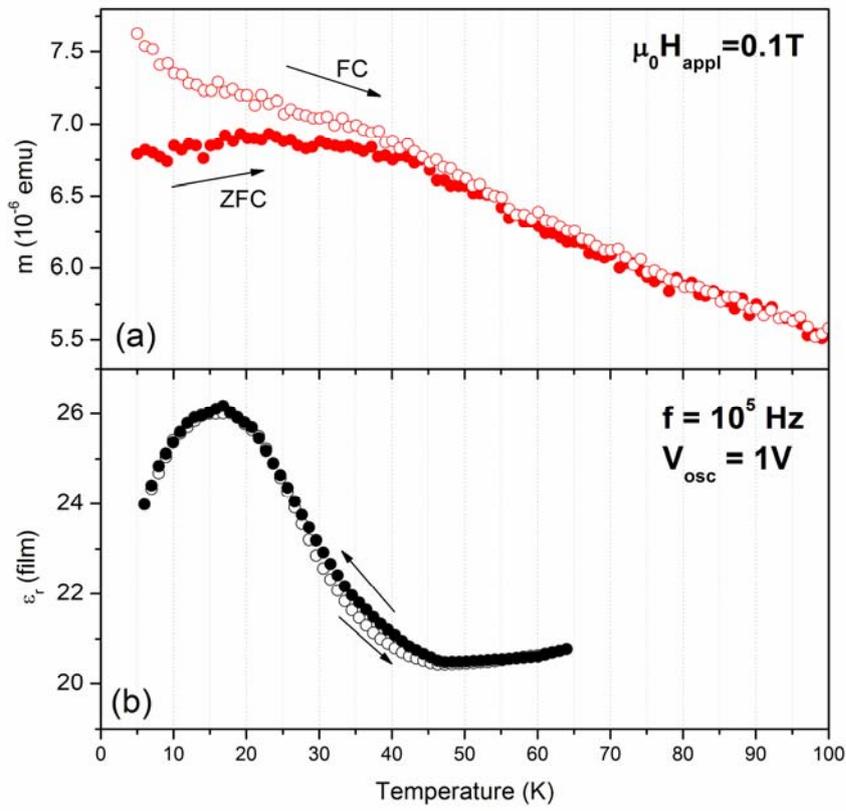

**Figure 4**

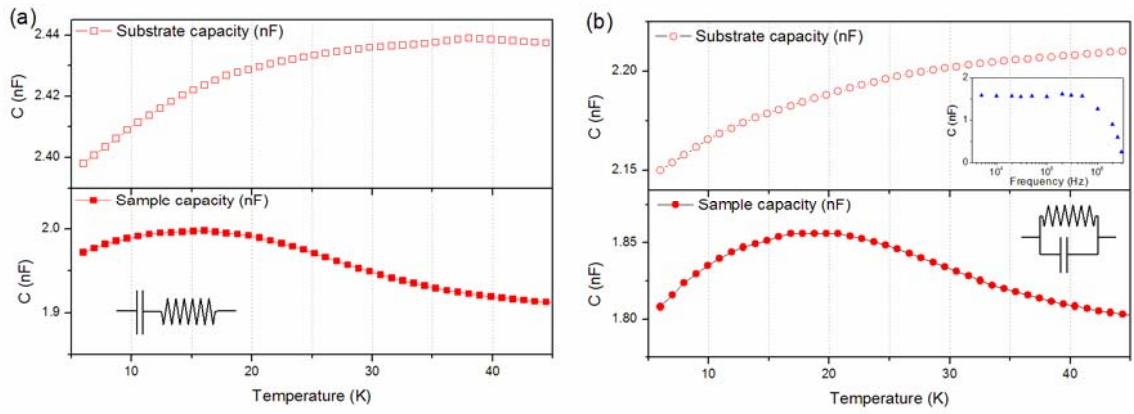